# Facts and Figuring: An Experimental Investigation of Network Structure and Performance in Information and Solution Spaces

JESSE SHORE, Boston University
ETHAN BERNSTEIN, Harvard University
DAVID LAZER, Northeastern University

1. OVERVIEW

Knowledge work in organizations, by definition, leverages contributions of many individuals to solve problems too complex for a single individual to tackle alone (e.g., Argote, McEvily, and Reagans, 2003). In organizations, such problem solving occurs not just at the individual level but also at the organizational level, expertise therefore becomes the core organizational asset, and superior answers (rather than, for example, superior production) define organizational performance (Winch and Schneider, 1993; Nickerson and Zenger, 2004). The proliferation of knowledge work has produced a mandate to understand the conditions under which successful integration of intra-organizational knowledge will best enable advanced problem solving and higher performance. However, although much has been written about organizational knowledge and problem solving, our understanding of how to optimize problem solving in knowledge-based organizations remains limited.

In particular, the question of how *clustered* (Centola and Macy, 2007) organizational networks should be to optimize problem solving remains elusive in theory and practice. Existing research on problem solving in networks gives incongruent, seemingly contradictory answers to the question of how densely clustered network ties should be for optimal problem solving performance. As discussed in detail in our manuscript, field work in organizations and some laboratory experiments on human subjects have shown a clear positive relationship between clustering and networked problem solving success (e.g., McCubbins, et al., 2009; Enemark et al, 2011; Enemark, McCubbins, and Weller, 2012); meanwhile, other laboratory experiments and research based on computational simulations have found the opposite result (e.g., Lazer and Friedman, 2007; Mason and Watts, 2012).

We present evidence from a laboratory experiment that resolves the apparent conflict in existing literature. We observe that while problem solving requires both the act of *searching for information*, or the facts that may be important pieces of the puzzle, and the act of *searching for solutions*, or the theories that combine puzzle pieces into an answer, prior studies have not adequately captured both dimensions. We hypothesize that these differences in domain—whether previous studies instrumented a search of information space, solution space or, as in the real world, both—have been responsible for the inconsistent findings on the connection between structure and problem solving. We therefore adopt a novel, data-rich experimental platform with greater verisimilitude than any in the past, which was specifically designed to include search and sharing of both information and solutions.

In adopting a richer experimental platform, we are able to reconcile the seemingly contradictory findings of prior work on network structure and problem solving. We find that clustering promotes exploration through information space, but it inhibits exploration through solution space. Through the active communication of information, individuals in a connected cluster tend to be in possession of the same knowledge and be aware of each other's theories. On the one hand, they therefore do not tend to search for information that another individual in the cluster has already found, and the cluster collectively explores information space more efficiently, and extensively, than they would have if they were not all connected to one another. On the other hand, because they are also all aware of each other's theories, there is an added tendency to interpret that information in the same way, which results in less exploration of theory space.

The same network structure, therefore, can promote or inhibit knowledge diversity, depending on whether that knowledge consists of information, or interpretations of information. The implication is that 'good' communication structures may only be good for parts of the process of collective problem solving: structures that are good now may be bad later.

## 1.1   Experiment

We selected a whodunit protocol, much like a game of Clue® or Cluedo®, in which the task involved piecing together clues to "connect the dots." Rather than creating a platform entirely from scratch, we were invited to customize a platform developed by the United States Department of Defense's Command and Control Research Program called ELICIT (Experimental Laboratory for Investigating Collaboration, Information-sharing, and Trust), which had many of the characteristics we sought. While we modified much of the platform, we agreed to keep the nature of the Department of Defense's whodunit task: predicting the who, what, where, and when of an impending terrorist attack.

There was a discrete set of actions available to experimental subjects. Through the web-based interface, they could search for new clues by entering a keyword into a search text box and clicking a button (much like a common online search engine). They could share these clues with one or more of their neighbors and, if they wished, add free text annotations to these shared clues (like a common inbox). They could register their theories by typing them into the separate spaces given for the who, what, where and when sub-problems. Finally, they could check their neighbors' registered solutions.

Since the information space was large, participants would have been extremely unlikely to find all of the necessary clues to solve all of the problems within the time limit on their own. However, because participants could also choose to share clues with their network neighbors, annotate those clues, and view their neighbors' registered solutions, the task was a collective problem-solving situation. Participants received no explicit, separate incentive, positive or negative, for sharing.

We recruited 417 unique individuals, who played a total of 1120 person-rounds. We tested four 16-person network treatments (see Figure for visualizations). At the top left is the so-called "caveman" network (Watts, 1999), containing four four-person cliques.  The "hierarchy" is likewise composed of four such cliques, but arranged in a conventional centralized structure. The "rewired caveman" is a small world network, constructed by removing links from the caveman, then adding links that create shortcuts through the network.  As a result, individuals in the rewired caveman are "closer together" topologically. The rewired caveman is also more centralized and less clustered than the caveman. Finally, there is the ring "lattice," which is neither clustered nor centralized. 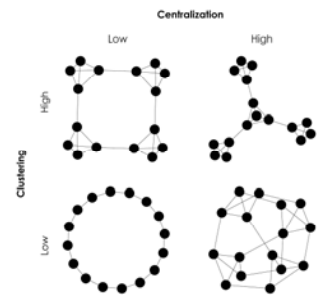

## 1.2   Brief Overview of Results

The most clustered networks, the caveman and the hierarchy, delivered the highest collective problem-solving performance in our Clue®-type problem-solving task. The more interesting insights, however, lie in the details behind that performance improvement.

Holding degree constant, increased clustering did not result in an increased number of searches. Instead, at the collective level, the facts found by the caveman network were significantly less redundant than those found by the ring lattice and the rewired caveman treatments. In other words, if the objective of information search was exploration, the caveman network was more efficient, in that they collectively covered more new ground with each search. Holding clustering constant, greater degree was actually associated with *decreased* number of searches through information space.

At the same time, exploration of theories, or what we call search of solution space, operated quite differently. At the collective level, the total number of unique theories registered was *highest* in the unclustered networks: the lattice and the rewired caveman both had significantly more unique theories than the caveman network with the hierarchy between the caveman and rewired cave. Clustering is associated with more checks of neighbors' theories and more outright copying of their theories. Consistent with the predictions of the information processing literature (Galbraith, 1974; Schneider, 1987), we also found that degree and total information received were correlated with less exploration (greater copying). At the collective level, the two clustered networks had significantly fewer unique theories registered in the aggregate than did the unclustered networks.

Ironically, the combination of these effects meant that clustering improved collective results without benefiting individuals themselves--no additional performance benefits accrued to individuals located in more highly clustered positions within a network.

## 1.3   Summary Implications

It is well established that network structure can influence problem solving performance (e.g., Grannovetter, 1973; Hansen, 1999; Burt, 2004), and yet a clear understanding of the role of clustering, a basic structural network variable, has remained elusive. By theoretically and experimentally disentangling information search from solution search -- two core domains of problem-solving -- this study indicates that the effect of clustering is opposite in those two domains. Clustering promotes exploration through information space, but depresses exploration through solution space. Whether increased clustering improves or impairs performance will depend on whether the immediate task or problem-solving stage benefits more from exploration of facts or, instead, the figuring that comes through the exploration of theories that interpret those facts.

Awareness of the differential performance effects of clustering for problem solving in information space from problem solving in solution space presents two challenges. For networks of problem-solving individuals, whether they represent groups, organizational units, whole organizations, or clusters of organizations, the challenge is one of leadership, such that leaders find ways of pairing the domain of the problem solving task, whether facts or figuring, with an appropriate network structure, whether clustered or not, to improve problem-solving performance. For scholars of both networks and information science, the challenge is one of further research: integrating our basic finding of the distinction between facts and figuring into the examination of how different network structures impact performance may help not only to resolve existing conflicts in disparate yet interconnected literatures, but also open up substantial opportunities for greater, coherent understanding of how we can set the conditions for problem-solving success in networks.

Clustering is a double-edged sword. It has the power to bring members of a network to generate more non-redundant information, but it also has the power to discourage theoretical exploration. Until one knows whether a problem-solving task involves searching for facts or searching for answers, it is impossible to predict the influence of clustering on organizational performance.